\documentclass{aipproc}
\usepackage{epsf}  
\layoutstyle{6x9}
\begin{document}
\addtolength{\topmargin}{+50pt}
%....................................................................

%............................ definitions/newcommands ...............

%\def\beq{\begin{eqnarray}}
\newcommand{\beq}{\begin{eqnarray}}
\newcommand{\eeq}{\end{eqnarray}}

%....................................... Various boldface symbols

%\def\s{\mbox{\boldmath$\displaystyle\mathbf{\sigma}$}}
\newcommand{\s}{\bm{\sigma}}

\newcommand{\be}{\bm{\eta}}

\newcommand{\bp}{\bm{\pi}}
\newcommand{\J}{\bm{J}}

\newcommand{\K}{\mathbf{K}}

\newcommand{\cp}{\bm{P}}

\newcommand{\p}{\bm{p}}

\newcommand{\hp}{\bm{\widehat{\p}}}

\newcommand{\x}{\bm{x}}

\newcommand{\0}{\bm{0}}

\newcommand{\bv}{\bm{\varphi}}

\newcommand{\bx}{\bm{\xi}}

\newcommand{\bs}{\bm{\sigma}}

\newcommand{\bc}{\bm{\chi}}

\newcommand{\hbv}{\bm{\widehat\varphi}}

\newcommand{\hbxi}{\bm{\widehat\xi}}

\newcommand{\bn}{\bm{\nabla}}

\newcommand{\bl}{\bm{\lambda}}

\newcommand{\br}{\bm{\rho}}

\newcommand{\ar}{\stackrel{\hspace{0.04truecm}grav. }
{\mbox{$\longrightarrow$}}}

%......................................................................

%......................................................................

%..................................................................

\title{Spin and Resonant States in QCD}

\author{M. Kirchbach}
{address={Instituto de F\'{\i}sica, 
         Universidad Aut\'onoma de San Luis Potos\'{\i},
         Av. Manuel Nava 6, San Luis Potos\'{\i}, S.L.P. 78240, M\'exico}}

\begin{abstract}
I make the case that the nucleon
excitations do not exist as isolated higher spin states but are fully
absorbed by $\left({K\over 2},{K\over 2}\right)\otimes
{\Big[}\left({1\over 2},0\right)\oplus \left(0,{1\over 2}\right)   {\Big]}$
multiplets taking their origin from the rotational and
vibrational excitations of an underlying quark--diquark string.
The $\Delta (1232)$ spectrum presents itself as
the exact replica (up to $\Delta (1600)$) 
of the nucleon spectrum with the $K$- clusters being 
shifted upward by about 200 MeV.
QCD inspired arguments support legitimacy of the quark-diquark 
string. 
The above $K$ multiplets can be mapped (up to form-factors) onto Lorentz 
group representation spaces of the type  $\psi_{\mu_1...\mu_K}$,
thus guaranteeing  covariant description of resonant states.
The quantum $\psi_{\mu_1...\mu_K}$ states are of
multiple spins at rest, and of undetermined spins elsewhere.
\end{abstract}

%\pacs{11.30Cp, 11.30Hv, 11.30Rd, 11.20Gk}

\maketitle

%............................ definitions ...............
\def\beq{\begin{eqnarray}}
\def\eeq{\end{eqnarray}}

\def\A{{\mathcal A}^\mu}
\def\W{{\mathcal W}_\mu}
%.......................

%............................ definitions ...............
\def\beq{\begin{eqnarray}}
\def\eeq{\end{eqnarray}}

%....................................... Various boldface symbols

\def\s{\mbox{\boldmath$\displaystyle\mathbf{\sigma}$}}
\def\S{\mbox{\boldmath$\displaystyle\mathbf{\Sigma}$}}
\def\J{\mbox{\boldmath$\displaystyle\mathbf{J}$}}
\def\K{\mbox{\boldmath$\displaystyle\mathbf{K}$}}
\def\P{\mbox{\boldmath$\displaystyle\mathbf{P}$}}
\def\p{\mbox{\boldmath$\displaystyle\mathbf{p}$}}
\def\hp{\mbox{\boldmath$\displaystyle\mathbf{\widehat{\p}}$}}
\def\x{\mbox{\boldmath$\displaystyle\mathbf{x}$}}
\def\0{\mbox{\boldmath$\displaystyle\mathbf{0}$}}
\def\bv{\mbox{\boldmath$\displaystyle\mathbf{\varphi}$}}
\def\hbv{\mbox{\boldmath$\displaystyle\mathbf{\widehat\varphi}$}}

\def\bl{\mbox{\boldmath$\displaystyle\mathbf{\lambda}$}}
\def\bl{\mbox{\boldmath$\displaystyle\mathbf{\lambda}$}}
\def\br{\mbox{\boldmath$\displaystyle\mathbf{\rho}$}}
\def\1{\openone}
\def\bfhh{\mbox{\boldmath$\displaystyle\mathbf{(1/2,0)\oplus(0,1/2)}\,\,$}}

\def\mn{\mbox{\boldmath$\displaystyle\mathbf{\nu}$}}
\def\amn{\mbox{\boldmath$\displaystyle\mathbf{\overline{\nu}}$}}

\def\mne{\mbox{\boldmath$\displaystyle\mathbf{\nu_e}$}}
\def\amne{\mbox{\boldmath$\displaystyle\mathbf{\overline{\nu}_e}$}}
\def\rlh{\mbox{\boldmath$\displaystyle\mathbf{\rightleftharpoons}$}}

\def\wm{\mbox{\boldmath$\displaystyle\mathbf{W^-}$}}
\def\hh{\mbox{\boldmath$\displaystyle\mathbf{(1/2,1/2)}$}}
\def\h00h{\mbox{\boldmath$\displaystyle\mathbf{(1/2,0)\oplus(0,1/2)}$}}
\def\znbb{\mbox{\boldmath$\displaystyle\mathbf{0\nu \beta\beta}$}}
%.......................

%........................................................

%........................................................

\section{Spectra of light-quark baryons}
Understanding the spectrum of the most simplest composite systems has always
been a key point in the theories of the micro-world. Recall that quantum
mechanics was established only after the successful description of the 
experimentally observed regularity patterns (such like the Balmer- series)
in the excitations of the hydrogen atom. Also in solid state physics, the 
structure of the low--lying excitations, be them without or with a gap,
has been decisive for unveiling the dynamical properties of
the many-body system-- ferromagnet versus superconductor, and the
relevant degrees of freedom, magnons versus Cooper pairs.
In a similar way, the regularity patterns of the nucleon excitations
are decisive for uncovering the relevant subnucleonic degrees of freedom 
and the dynamical properties of the theory of strong interaction-- 
the Quantum Chromo- Dynamics.

Despite its long history, amazingly, the structure of the nucleon spectrum
is far from being settled. This is due to the fact that the first facility
that measured nucleon levels, the Los Alamos Meson Physics Facility 
(LAMPF) failed to find all the states that were predicted by the excitations 
of three quarks. Later on, the Thomas Jefferson National Accelerator
Facility
(TJNAF) was designed to search (among others) for those 
``missing resonances''.
At present, all data have been collected and are awaiting evaluation
\cite{Burkert}.  

In a series of papers \cite{MK97-98} a new and subversive look on the
reported data in Ref.~\cite{PART} was undertaken. There I drew attention 
to the ``{\tt Come--Together}'' of resonances of different spins 
and  parities to narrow mass bands in the nucleon spectrum and, its exact
replica in the  $\Delta (1232) $ spectrum (see Fig.~ 1). 

The first group of states consists of two spin-${1\over 2}$
states of opposite parities and a single spin-${3\over 2}^-$.
The second group has three parity degenerate states with spins
varying from ${1\over 2}^\pm$ to ${5\over 2}^\pm $, and a 
single spin-${7\over 2}^+$ state.
Finally, the third group has five parity degenerate states with spins ranging
from ${1\over 2}^\pm$ to ${9\over 2}^\pm $, and a single spin 
${{11}\over 2}^+$ state 
(see Ref.~\cite{MK_Evora} for the complete $N$ and $\Delta (1232) $ spectra).
A comparison between the $N$ and $\Delta (1232) $ spectra shows that they
are identical up to two ``missing'' resonances
on the nucleon side (these are the counterparts of the $F_{37}$ and $H_{3, 11}$
states of the $\Delta $ excitations) and up to three ``missing'' states on 
the $\Delta $ side (these are the counterparts of the nucleon 
$P_{11}$, $P_{13}$, and $D_{13}$ states from the third group). 
The $\Delta (1600)$ resonance
which is most probably and independent hybrid state, is the only state
that at present seems to drop out of our systematics.

The existence of identical nucleon- and $\Delta $ crops of resonances 
raises the question as to what extent are we facing here a new type of 
symmetry which was not anticipated by any model or theory before.
The next section devotes itself to answering this question.

\section{Quark--diquark string excitations}
 
Baryons in the quark model are considered as constituted of three
quarks in a color singlet state. It appears naturally, therefore,
to undertake an attempt of describing the baryonic system
by means of algebraic models developed for
the purposes of triatomic molecules,
a path already pursued by  Refs.~\cite{U(7)}.

In the dynamical limit $U(7)\longrightarrow U(3)\times U(4)$
of the three quark system, two of the quarks reveal a stronger 
pair correlation to a diquark (Dq), 
while the third quark (q) acts as a spectator.
The diquark approximation~\cite{Torino} turned out to be rather convenient
in particular in describing various properties of the ground state 
baryons \cite{ReinA}, \cite{Kusaka}. Within the context of the 
quark--diquark (q-Dq) 
model, the ideas of the rovibron model, known from the spectroscopy of 
diatomic molecules \cite{diat}, can be applied to the description
of the rotational-vibrational (rovibron) excitations of the q--Dq system.

\noindent
\underline{{\it Rovibron model for the quark--diquark system.}}
In the rovibron model (RVM) the relative q--Dq motion
is described by means of four types of boson creation
operators $s^+, p^+_1, p^+_0$, and $p^+_{-1}$. 
The operators $s^+$ and $p^+_m$ in turn 
transform as rank-0, and rank-1 spherical tensors,
i.e. the magnetic quantum number $m$ takes
in turn the values $m=1$, $0$, and $-1$.
In order to construct boson-annihilation operators that
also transform as spherical tensors, one introduces
the four operators $\widetilde{s}=s$, and
$\widetilde{p}_m=(-1)^m\, p_{-m}$.
Constructing rank-$k$ tensor product of
any rank-$k_1$ and rank-$k_2$ tensors, say, $A^{k_1}_{m_1}$ 
and $A^{k_2}_{m_2}$, is standard and given by
\begin{equation}
\lbrack A^{k_1}\otimes A^{k_2}\rbrack^k_m =
\sum_{m_1,m_2}(k_1m_1 k_2m_2\vert km)\, A^{k_1}_{m_1}A^{k_2}_{m_2}\, .
\label{clebsh}
\end{equation}
Here, $(k_1m_1k_2m_2\vert km)$ are the standard $O(3)$ Clebsch-Gordan
coefficients.

Now, the lowest states of the two-body system are identified with $N $
boson states and are characterized by the ket-vectors 
$\vert n_s\, n_p\, l\, m\rangle $ (or, a linear combination of them)
within a properly defined Fock space. The constant  
$N=n_s +n_p$ stands for the total number of $s$- and $p$ bosons
and plays the r\'ole  of a parameter of the theory.
In molecular physics, the parameter $N$ is usually associated
with the number of molecular bound states.
The group symmetry of the rovibron model is well known to be $U(4)$. 
The fifteen generators of the associated $su(4)$ algebra 
are determined as the following set of bilinears 
\begin{eqnarray}
A_{00}=s^+ \widetilde{s}\, , &\quad& A_{0m}=s^+\widetilde{p}_m\, , 
\nonumber\\
A_{m0}=p^+_m\widetilde{s}\, , &\quad & A_{mm'}=p^+ _m
\widetilde{p}_{m'}\, .
\label{RVM_u4}
\end{eqnarray} 
The $u(4)$ algebra is then recovered by the following 
commutation relations
\begin{equation}
\lbrack A_{\alpha\beta},A_{\gamma\delta}\rbrack_-=
\delta_{\beta \gamma}A_{\alpha\delta}-
\delta_{\alpha\delta}A_{\gamma\beta}\, .
\end{equation}
The operators associated with physical observables can then be expressed
as combinations of the $u(4)$ generators.
To be specific, the three-dimensional angular momentum takes the form
\begin{equation}
L_m=\sqrt{2}\, \lbrack p^+ \otimes \widetilde{p}\rbrack^1_m \, .
\label{a_m}
\end{equation}
Further operators are  $(D_m$)-- and $(D'_m$) defined as 
\begin{eqnarray}
D_m &=&\lbrack p^+\otimes \widetilde{s}+s^+\otimes 
\widetilde{p}\rbrack^1_m\, ,\\
\label{x_dipol_rvm}
D_m '&=&i\lbrack p^+\otimes \widetilde{s}-s^+\otimes 
\widetilde{p}\rbrack^1_m\, ,
\label{p_dipol_rvm}
\end{eqnarray}
respectively.
Here, $\vec{D}\, $ plays the r\'ole of
the electric dipole operator.

{}Finally, a quadrupole operator $Q_m$ can be constructed as
\begin{equation}
Q_m=\lbrack p^+\otimes \widetilde{p}\rbrack^2_m\, ,
\quad \mbox{with}\quad m=-2,..., +2\, .
\label{quadr_rvm}
\end{equation}
The $u(4)$ algebra has the two algebras $su(3)$, and $so(4)$, as 
respective sub-algebras. The $so(4)$ sub-algebra of interest here,
is constituted by the three components of the angular momentum operator 
$L_m$, on the one side,
and the three components of the operator $D_m'\, $, on the other side. 
The chain of reducing $U(4)$ down to $O(3)$  
\begin{equation}
U(4)\supset O(4)\supset O(3)\, ,
\label{chains}
\end{equation}
corresponds to an exactly soluble RVM limit.
The Hamiltonian of the RVM in this case
is  constructed as a properly chosen function of the Casimir 
operators of the algebras of the subgroups entering the chain. 
{}For example, in case one approaches $O(3)$ via $O(4)$,
the Hamiltonian of a dynamical $SO(4)$ symmetry can be cast into 
the form \cite{KiMoSmi}:
\begin{equation}
H_{RVM}=H_0 - f_1\, \left(4 {\cal C}_2\left( so(4)\right) +1\right)^{-1}
+f_2\, {\cal C}_2(so(4) )\, .
\label{H_QRVM}
\end{equation}
The Casimir operator ${\cal C}_2\left(so(4)\right)$ is defined accordingly as
\begin{equation}
{\cal C}_2\left( so(4)\right)={1\over 4}\left( \vec{L}\, ^2 + 
\vec{D}\, ' \, ^2
\right)\, 
\label{so(4)_Casimir}
\end{equation}
and has an eigenvalue of ${K\over 2}\left( {K\over 2}+1 \right)$.
Here, the parameter set has been chosen as  
\begin{equation}
H_0= M_{N/\Delta } +f_1\, ,\quad  f_1=600\,\, \mbox{MeV}\, , \quad 
f_2^N=70\, \, \mbox{MeV}\, , \quad
f_2^\Delta =40 \, \, \mbox{MeV}\, .
\label{f_s}
\end{equation}
Thus, the $SO(4)$ dynamical symmetry limit of the
RVM picture of baryon structure motivates  
existence of quasi-degenerate resonances gathering to crops 
in both the nucleon- and $\Delta $ baryon spectra.
The Hamiltonian that will fit masses of the reported
cluster states is exactly the one in Eq.~(\ref{H_QRVM}).

In order to demonstrate how the RVM applies to baryon spectroscopy,
let us consider the case of q-Dq states associated with $N=5$
and for the case of a $SO(4)$ dynamical symmetry. 
It is of common
knowledge that the totally symmetric irreps of the $u(4)$ algebra 
with the Young scheme $\lbrack N\rbrack$ contain the 
$SO(4)$ irreps $\left({K\over 2}, {K\over 2}\right)$
(here $K$ plays the role of the four-dimensional angular momentum) 
with
\begin{equation}
K=N, N-2, ..., 1 \quad \mbox{or}\quad 0\, .
\label{Sprung_K}
\end{equation}
Each one of the $K$- irreps contains $SO(3)$ multiplets with
three dimensional angular momentum
\begin{equation}
l=K, K-1, K-2, ..., 1, 0\, .
\label{O3_states}
\end{equation}
In applying the branching rules in Eqs.~(\ref{Sprung_K}),
(\ref{O3_states})
to the case $N=5$, one encounters the series of levels
\begin{eqnarray}
K&=&1: \quad l=0,1;\nonumber\\
K&=&3: \quad l=0,1,2,3;\nonumber\\
K&=&5: \quad l=0,1,2,3,4,5\, .
\label{Ns_Ks}
\end{eqnarray}
The parity carried by these levels is $\eta (-1)^{l}$ where
$\eta $ is the parity of the relevant vacuum. In coupling now the
angular momentum in Eq.~(\ref{Ns_Ks}) to the spin-${1\over 2}$ of the three
quarks in the nucleon, the following sequence of states is obtained:
\begin{eqnarray}
K&=&1: \quad \eta J^\pi={1\over 2}^+,{1\over 2}^-, {3\over 2}^-\, ;
\nonumber\\
K&=&3: 
\quad \eta J^\pi={1\over 2}^+,{1\over 2}^-, {3\over 2}^-,
{3\over 2}^+, {5\over 2}^+, {5\over 2}^-, {7\over 2}^- \, ;
\nonumber\\
K&=&5: \quad \eta J^\pi={1\over 2}^+,{1\over 2}^-, {3\over 2}^-,
{3\over 2}^+, {5\over 2}^+, {5\over 2}^-, {7\over 2}^- ,
{7\over 2}^+, {9\over 2}^-, {11\over 2}^-\, .
\label{set_q}
\end{eqnarray}
Therefore, rovibron states of half-integer spin transform according to  
$\left( {K\over 2},{K\over 2}\right) \otimes \left[
\left({1\over 2},0 \right) \oplus 
\left( 0,{1\over 2} \right)\right] $
representations of $SO(4)$.
The isospin structure is accounted for pragmatically through
attaching to the K--clusters an isospin spinor
$\chi^I$ with $I$ taking the values $I={1\over 2}$ 
and $I={3\over 2}$ for the nucleon, and the $\Delta $ states,
respectively.
As illustrated by Fig.~1, the above quantum numbers cover
both the nucleon and the $\Delta $ excitations.
\begin{figure}[htb]
\leavevmode
\epsfxsize=7cm
\epsffile{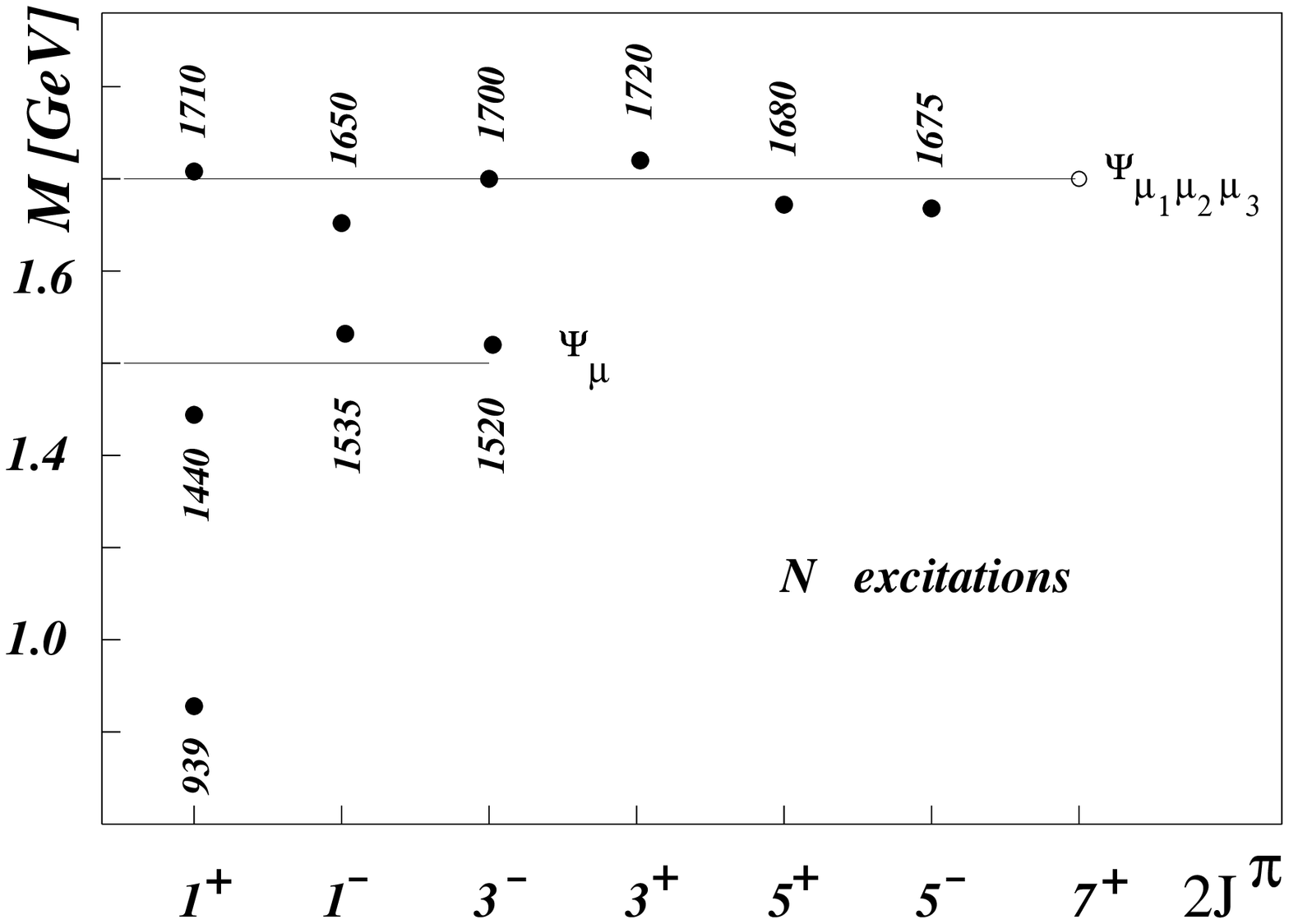}
\epsfxsize=7cm
\epsffile{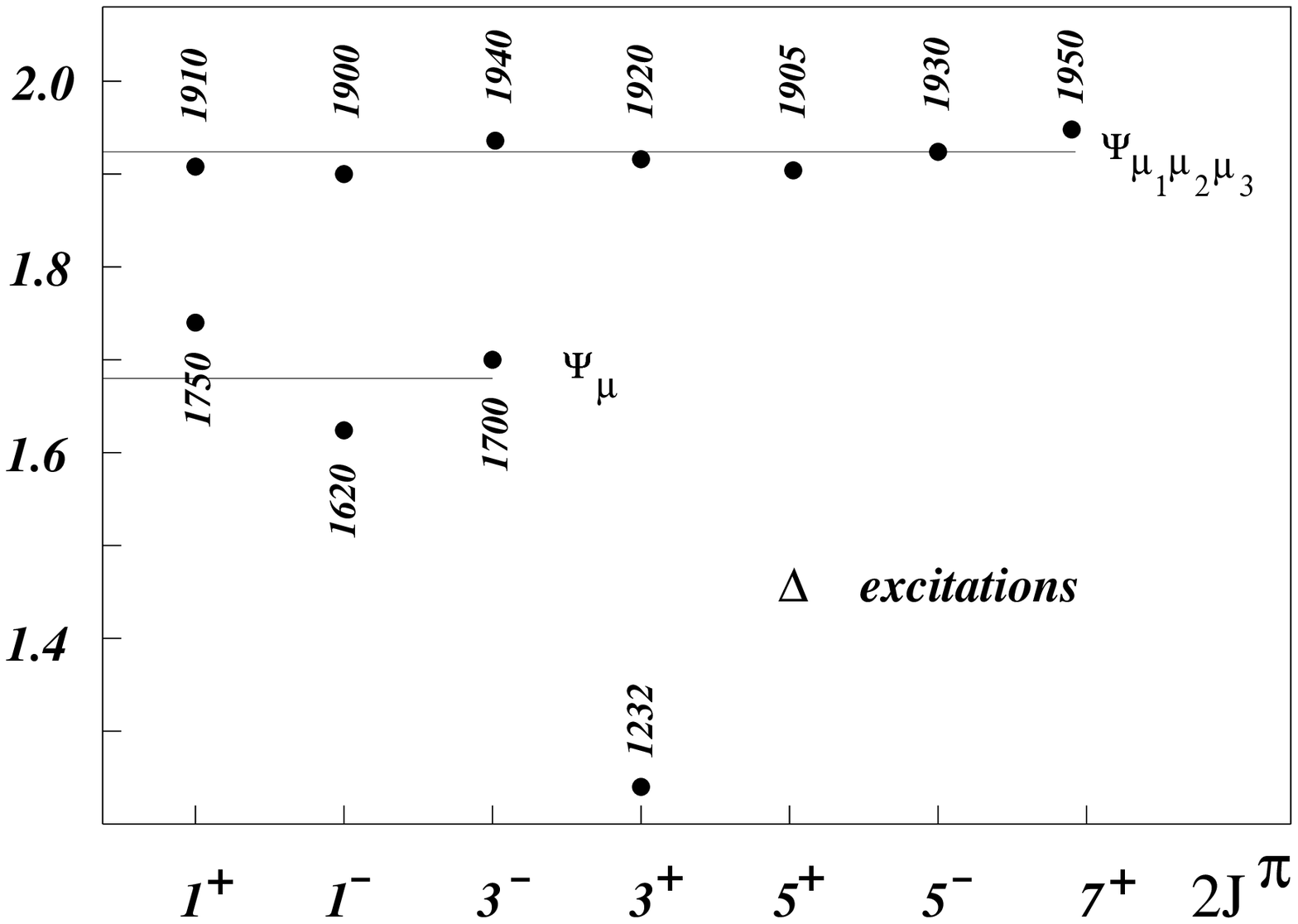}
\caption{
Summary of the data on the nucleon and the $\Delta$ resonances.
The breaking of the mass degeneracy for each of the clusters
at about $5\%$ may in fact be an artifact of the data analysis,
as has been suggested by H\"ohler \cite{gh}.
The filled circles represent known resonances, while 
the sole empty circle corresponds to a prediction.
Figure taken from \cite{PLB_02}. 
}
\label{kut}
\end{figure}
The states in  Eq.~(\ref{set_q}) are degenerate and the dynamical symmetry
is $SO(4)$. 

\noindent
\underline{{\it  Observed and ``missing'' resonance clusters within the 
rovibron model.}}
The comparison of the states in Eq.~(\ref{set_q}) with the reported ones
in  Fig.~1 shows that the predicted sets are in agreement 
with the characteristics of the non-strange baryon excitations with
masses below $\sim $ 2500 MeV, provided, the parity $\eta $ of the vacuum
changes from scalar ($\eta =1$) for the $K=1$, to pseudoscalar 
($\eta =-1$) for the $K=3,5$ clusters.
A pseudoscalar ``vacuum'' can be modeled in terms of
an excited  composite  diquark carrying an internal 
angular momentum  $L=1^-$ and maximal spin $S=1$. In one of
the possibilities the total spin of such
a system can be $\vert L-S\vert = 0^-$.
To explain the properties of the ground state, one has to
consider separately even $N$ values, such as, say, $N'=4$.
In that case another branch of excitations, with $K=4$, $2$, and
$0$ will emerge. The $K=0$ value characterizes the 
ground state,  $K=2$ corresponds to
$\left( 1,1\right)\otimes 
\lbrack\left( {1\over 2},0\right)\oplus 
\left(0,{1\over 2}\right) \rbrack $, while  $K=4$ corresponds to
$\left( 2,2\right) \otimes\lbrack\left( {1\over 2},0\right)\oplus 
\left(0,{1\over 2}\right) \rbrack $. 
These are the multiplets that we 
will associate with the  ``missing'' resonances  
predicted by the rovibron model.
In this manner, reported and ``missing'' resonances  fall apart 
and populate  distinct $U(4)$- and $SO(4)$ representations.
In making observed and ``missing'' resonances distinguishable,
reasons for their absence or, presence in the spectra are
easier to be searched for.
In accordance with Ref.~\cite{MK_2000}
we here will treat the $N=4$ states to be all of natural parities
and identify them with the nucleon $(K=0)$, the natural parity $K=2$,
and  the natural parity $K=4$--clusters.
We shall refer to the latter as `missing'' rovibron  clusters.
In Table I we list the masses of the K--clusters concluded from
Eqs.~(\ref{H_QRVM}), and (\ref{f_s}).
\begin{table}[htbp]
\caption{
Predicted mass distribution of observed (obs), and
missing (miss) rovibron  clusters (in MeV) according to
Eqs.~(9,11). 
The sign of $\eta $ in Eq.~(15) determines natural-
($\eta =+1$), or, unnatural ( $\eta =-1$) parity states.
%{}All $\Delta $ excitations have been calculated
%with  $m_2=40$ MeV rather than
%with the nucleon value of $m_2=70$ MeV. 
The experimental mass averages of the resonances from a given
K--cluster have been labeled by ``exp''.
%The nucleon and $\Delta $ ground state masses  
%$M_N$ and $M_\Delta $ were taken to equal their experimental values. 
}
\vspace*{0.21truein}
\begin{tabular}{lccccccc}
\hline
~\\
K & sign $\eta $ & N$^{\mbox{obs} }$  & N$^{\mbox{exp}}$ &
 $ \Delta^{\mbox{obs}}$ & $\Delta^{\mbox{exp}}$ &  
   N$ ^{\mbox{miss}}$ & $\Delta^{\mbox{miss}}$  \\
\hline
~\\
0 & + &939 &939 & 1232& 1232 & &  \\
1&+  &1441 & 1498 & 1712 &1690 &         & \\
2&+  &     &      &      &     &  1612   & 1846 \\
3&-  &1764 &1689  & 1944 &1922 &         &     \\
4&+  &     &      &      &     &  1935   & 2048 \\
5&-  &2135 &2102  & 2165 &2276 &         &      \\
\hline
\end{tabular}
\end{table}

\noindent
\underline{{\it Spin and quark--diquark in QCD.}}
The necessity for having a quark--diquark configuration within the
nucleon follows directly from QCD arguments. 
In Refs.~\cite{Doug1}, and \cite{Doug2} the notion of 
spin in QCD was re-visited in connection with the proton spin puzzle.
As it is well known, the spins of the valence quarks are by themselves not
sufficient to explain the spin-${1\over 2}$ of the nucleon.
Rather, one needs to account for
the orbital angular momentum of the quarks 
(here denoted by $L_{QCD}$) and the angular momentum carried
by the gluons (so called field angular momentum, $G_{QCD}$):
\begin{eqnarray*}
{1\over 2} &=&{1\over 2}\Delta \Sigma  +L_{QCD}  +G_{QCD} \, \\
= \int d^3 x {\Big[} {1\over 2}\bar \psi \vec{\gamma}\gamma_5\psi
&+&\psi^\dagger (\vec{x}\times (-\vec{D}))\psi 
+\vec{x}\times (\vec{E}^a\times \vec{B}^a){\Big]}\, .
\end{eqnarray*}
In so doing one encounters the problem that neither $L_{QCD}$, nor $G_{QCD}$
satisfy the spin $su(2)$ algebra. If at least  
$\left(L_{QCD} +G_{QCD}\right)$ is to do so,
\beq
{\Big[} 
\left( L_{QCD}^i +G^i_{QCD} \right),
\left( L_{QCD}^j +G^j_{QCD}\right) 
{\Big]}
=i
\epsilon^{ijk} \left( L_{QCD}^k +G^k_{QCD}\right)\, ,
\label{su2_alg}
\eeq	
then $\vec{E}^{i;a}$ has to be  {\it restricted}  
to a chromo-electric charge, while $\vec{B}^{i;a} $ 
has to be a chromo-magnetic dipole according to, 
\beq
E^{i;a}={{gx' \,  ^i} \over {r'\, ^3}}T^a\, ,\quad 
B^{ia}= ({ {3x^i x^l m^l}\over r^5 } -
{m^i\over r^3})T^a\, ,
\eeq
where  $x'\, ^i=x^i-R^i$.
The above color fields are the perturbative one-gluon 
approximation typical for a diquark-quark structure. 
The diquark and the quark are in
turn the sources of the color Coulomb field,
and the color magnetic dipole field.
In terms of color and flavor degrees of freedom,
the nucleon wave function indeed has the required quark--diquark
form $
\vert p_{\uparrow} \rangle =
{ \epsilon_{ijk} \over \sqrt{18} }\,
\lbrack u^+_{i \downarrow} d^+_{j\uparrow} -
u^+_{i\uparrow}d^+_{j\downarrow} \rbrack\,  u^+_{k\uparrow}\,\vert 0\rangle .
$
A similar situation appears when looking for covariant QCD solutions
in form of a membrane with the three open ends being associated with the 
valence quarks. When such a membrane stretches to a string, so that
a linear action (so called gonihedric string) can be used, one again
encounters that very $K$-cluster degeneracies in the excitations 
spectra of the baryons, this time as a part of an infinite tower of states.
The result was  reported by Savvidy in Ref.~\cite{Savvidy}.
Thus the covariant spin-description provides an independent argument
in favor of a dominant quark-diquark configuration in the structure
of the nucleon, while search for covariant  
resonant QCD solutions leads once again to infinite $K$-cluster 
towers. 
The quark-diquark internal structure of the baryon's ground states
is just the configuration, the excited mode of which is described
by the rovibron model and which is the source of
the $\left( {K\over 2},{K\over 2}\right)\otimes
\Big[  \left( {1\over 2},0\right)\oplus \left(0,{1\over 2}\right)
\Big] $ patterns.

In Ref.~\cite{KiMoSmi} we presented  the four dimensional Racah algebra that 
allows to calculate transition probabilities for electromagnetic 
de-excitations of the rovibron levels. The interested reader is invited 
to consult the quoted article for details. Here I restrict myself to 
reporting the following two results:
(i) All resonances from a $K$- mode have same widths.
(ii) As compared to the natural parity $K=1$ states,
the electromagnetic de-excitations of the unnatural parity
$K=3$ and $K=5$ rovibron states appear strongly suppressed.  
To illustrate our predictions I compiled in Table 2 below
data on experimentally observed total widths of resonances 
belonging to  $K=3$, and $K=5$.
\begin{table}[htbp]
\caption{ Reported widths of resonance clusters }
\begin{tabular}{lcc}
\hline
~\\
K&Resonance & width [in GeV]  \\
\hline
~\\
3 &$N\left( {1\over 2}^-;1650\right)$ & 0.15  \\
3 &$N\left( {1\over 2}^+;1710\right)$ & 0.10   \\
3 &$N\left( {3\over 2}^+;1720\right)$ & 0.15    \\
3 &$N\left( {3\over 2}^-;1700\right)$ & 0.15     \\
3 &$N\left( {5\over 2}^-;1675\right)$ & 0.15      \\
3 &$N\left( {5\over 2};^+1680\right)$ & 0.13       \\
~\\
\hline
5 &$N\left( {3\over 2}^+;1900\right)$ & 0.50\\
5 &$N\left( {5\over 2}^+;2000\right)$ &0.49\\
~\\\hline
\end{tabular}
\end{table}
The suppression of the electromagnetic de--excitation modes of unnatural
parity states to the nucleon (of natural parity)
is shown in Table 3. It is due to
the vanishing overlap between the scalar diquark  in the latter case, 
and the  pseudo-scalar one, in the former. Non-vanishing widths can signal
small admixtures from natural parity states of same spins
belonging to even $K$ number states from the ``missing'' resonances.
\begin{table}[htbp]
\caption{ Reported helicity amplitudes of resonances }
\begin{tabular}{lcccc}
\hline
~\\
K& parity of the spin-0 diquark  
&Resonance  &$A^p_{{1\over 2}}$ & $A_{{3\over 2}}^2$ 
[in $10^{-3}$GeV$^{-{1\over 2}}$]  \\
\hline
~\\
3 &- & $N\left( {1\over 2}^+;1710\right)$  &  9 $\pm$22 &  \\
3 &- & $N\left( {3\over 2}^+;1720\right)$  &  18$\pm$30 &-19$\pm$20   \\
3 &- & $N\left( {3\over 2}^-;1700\right)$  & -18$\pm$30 & -2$\pm$24    \\
3 &- & $N\left( {5\over 2}^-;1675\right)$  & 19 $\pm$8  & 15$\pm$9      \\
3 &- & $N\left( {5\over 2};^+1680\right)$  & -15$\pm$6  &133$\pm$12      \\
~\\
\hline
1 &+ &$N\left( {3\over 2}^-;1520\right) $ & -24$\pm$9 & 166$\pm$ 5\\
~\\\hline
\end{tabular}
\end{table}
For example, the significant value of $A_{{3\over 2}}^p$ for 
$N\left( {5\over 2}^+;1680\right)$
from $K=3$ may appear as an effect of mixing with the 
N$\left( {5\over 2}^+;1612\right)$ state
from the natural parity ``missing'' cluster with $K=2$.
This gives one the idea to
use helicity amplitudes to extract ``missing'' states.

The above considerations show that a $K$-mode of an
excited  quark-diquark string (be the diquark scalar, or, pseudoscalar)
represents an independent entity (particle?) in its own rights which 
deserves its own 
name. To me the different spin facets of the $K$--cluster pointing into
different ``parity directions'' as displayed in Fig.~2 look like
barbs. That's why I suggest to refer to the $K$-clusters
as {\it barbed\/} states to emphasize the aspect of alternating parity. 
Barbs could also be associated with thorns 
(Spanish, espino), and {\it espinons\/} could be
another sound name for K-clusters.

\begin{figure}[htb]
\leavevmode
\epsfxsize=10cm
\epsffile{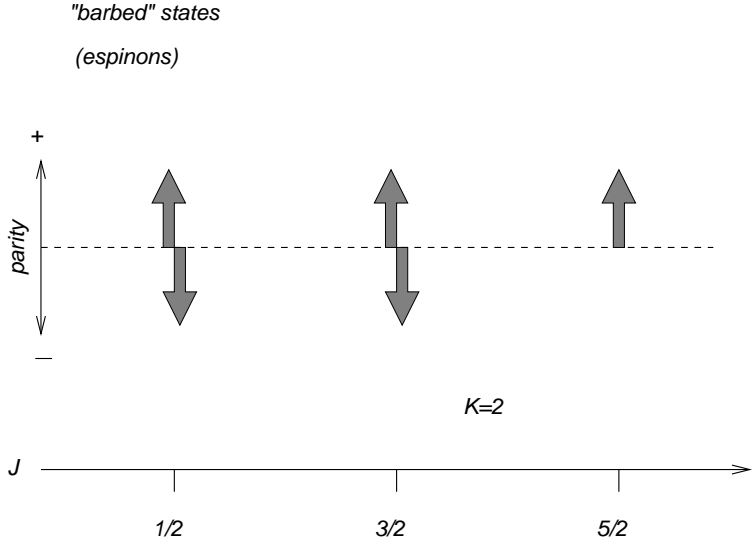}
\caption{
$K$-excitation mode of a quark-diquark string: {\it barbed \/} states 
({\it espinons\/} ).}
\label{esp}
\end{figure}

\section{Conclusions}
We argued that the {\tt Come-Together} of several parity degenerate 
states of increasing spins to $\psi_{\mu_1...\mu_K}$ multiplets
can be explained through  rotational-vibrational modes of an 
excited quark-diquark string, be the diquark scalar 
(in the respective observed $\psi_\mu$, and the ``missing'' 
$\psi_{\mu_1\mu_2}$, and $\psi_{\mu_1...\mu_4}$), or pseudoscalar 
(in the observed $\psi_{\mu_1...\mu_3}$, and $\psi_{\mu_1...\mu_5}$, 
respectively).
Each $K$ state consists
of $K$ parity couples and a single unpaired ``has--been''  
spin-$J=K+{1\over 2}$ at rest. 
The parity couples should not be confused with parity doublets.
The latter refer to states of equal spins, residing in distinct
Fock spaces built on top of opposite parity (scalar, and pseudoscalar)
vacua with  $\Delta l=0$ .
The parity degeneracy observed in the baryon spectra is an artifact
of the belonging of resonances to 
$\left({K\over 2},{K\over 2}\right)\otimes 
{\Big[}\left({1\over2},0\right)\oplus \left(0,{1\over 2}\right){\Big]}$,
in which case the opposite parities of equal spins originate
from underlying angular momenta differing by one unit, i.e. from
$\Delta l=1$. 
Chiral symmetry realization within the $K$-cluster scenario
means having coexisting scalar and pseudoscalar diquarks
(''vacua''), and consequently coexisting $K$-clusters  of both
natural and unnatural parities.   
Stated differently, if TJNAF is to supplement the unnatural
parity LAMPF ``espinons'' with $K$=3, and 5 by the natural parity $K$=2,4
ones, then we will have manifest mode of chiral symmetry
in the baryonic spectra.  
The total number of ordinary-spin states in our scenario
needs not be multiple of two, as it should be in case of parity doubling.

\section*{Acknowledgments}
Marcos Moshinsky and Yuri Smirnov brought their
unique expertise into uncovering the quark-diquark dynamics
behind the resonance clusters. 
Work partly supported by Consejo Nacional de Ciencia y
Tecnolog\'ia (CONACyT, Mexico) under grant number 32067-E.

\noindent
Special thanks to the organizers for having managed such a
memorable event.
Congratulations to Arnulfo Zepeda and Augusto Garcia
to their anniversaries and exemplary careers.

\end{document}